# Thermal and charge transport characteristics and fine details of the crystal structure in dodecaborides Lu$^N$B$_{12}$ (N = 10, 11, nat) with the Jahn-Teller instability

Nadezhda B. Bolotina [1,*], Alexander P. Dudka [1,2], Olga N. Khrykina [1,2], Vladimir V. Glushkov [2,3], Andrey N. Azarevich [2], Vladimir N. Krasnorussky [2], Slavomir Gabani [4], Natalya Yu. Shitsevalova [5], A.V. Dukhnenko [5], Volodymyr B. Filipov [5], Nikolay E. Sluchanko [2,3]

[1]Shubnikov Institute of Crystallography of Federal Scientific Research Centre 'Crystallography and Photonics' of Russian Academy of Sciences, 59 Leninskii prospect, 119333 Moscow, Russia

[2]Prokhorov General Physics Institute, Russian Academy of Sciences, 38 Vavilov Str., 119991 Moscow, Russia

[3]Moscow Institute of Physics and Technology (State University), 9 Institutskiy Per., 141700 Dolgoprudny, Russia

[4]Institute of Experimental Physics SAS, Watsonova 47, 04001 Košice, Slovakia

[5]Frantsevich Institute for Problems of Materials Science, National Academy of Sciences of Ukraine, 3 Krzhyzhanovsky Str., 03680 Kiev, Ukraine

Structure differences of isotopically different dodecaborides Lu$^N$B$_{12}$ (N = 10, 11, natural) and their impact on thermal and charge transport characteristics of the crystals have been first discovered. Atomic displacement parameters (ADPs) of Lu and B atoms are described in terms of the Einstein and Debye models, respectively. Characteristic Einstein and Debye temperatures are calculated directly from the x-ray data and corresponding ADPs are separated into temperature dependent and temperature independent components. The first component is a measure of thermal atomic vibrations whereas the second one is a sum of zero vibrations and static shifts of some atoms from their crystallographic positions. Such a local disordering is more expressed in Lu$^{nat}$B$_{12}$ with $^{10}$B : $^{11}$B ≈ 1 : 4 judging both from the large static ADP components and the Schottky anomalies in the heat capacity. Crystal structures are refined in $Fm\bar{3}m$ group but certain distortions of the ideal cubic unit-cell values are observed in all three crystals under study due to cooperative Jahn-Teller effect. The distortions are mainly trigonal or mainly tetragonal depending on the isotope composition. Low symmetry distribution of electron density reveals itself in Lu$^{nat}$B$_{12}$ in the form of the dynamic charge stripes oriented in selected directions close to some of <110>. The large static ADP components of Lu$^{nat}$B$_{12}$ are surprisingly combined with high conductivity of the crystal. One may suppose the static shifts (defects) are centers of pinning facilitating formation of additional conductive channels.

# I. INTRODUCTION

Dodecaborides $R$B$_{12}$ ($R$ = Y, Zr, Tb–Lu) attract attention of researchers as model objects for basic research in the field of solid state physics as well as perspective materials for practical applications.[1–4] Mechanical and thermal properties of these crystals, such as high melting temperatures, hardness, heat capacity, chemical stability, are mainly defined by boron framework built from empty cuboctahedra B$_{12}$ and large truncated octahedra B$_{24}$ centered by rare-earth or transition metals. It is generally known that almost all dodecaborides, including LuB$_{12}$, crystallize in cubic $Fm\bar{3}m$ group of symmetry similarly to NaCl but with Lu instead of Na and B$_{12}$ instead of Cl (Fig. 1).

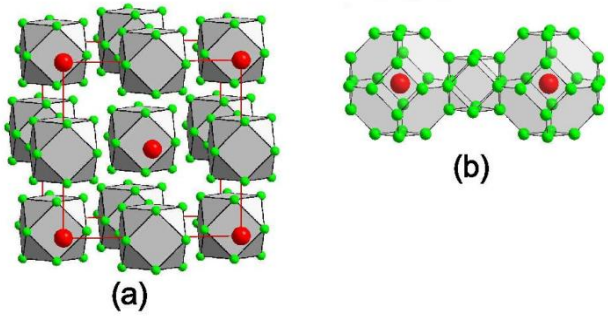

FIG. 1. (Color online) (a) a NaCl-like unit cell of LuB$_{12}$, with Lu (red spheres) as Na and B$_{12}$ clusters (B atoms shown as green spheres) as Cl. (b) Two B$_{24}$ polyhedra centered by Lu atoms and empty B$_{12}$ cuboctahedron between them.

It should be noted that measured properties of dodecaborides not always correspond to the ideal cubic structure. It has been established, in particular, that Raman spectra of LuB$_{12}$ and ZrB$_{12}$ contain modes prohibited by selection rules if static displacements of metallic cations from centrosymmetric 4$a$ position of the $Fm\bar{3}m$ group are not presumed. Structure disordering was confirmed using x-ray data and a small tetragonal distortion of cubic lattice was earlier observed at low temperatures.[5] Cooperative dynamic Jahn-Teller (JT) effect is assumed to be a cause of the boron framework deformation (Ref. 3) what makes an impact on crystal properties keeping in mind that the conduction band of LuB$_{12}$ is built on hybridized electronic states of Lu (5$d$) and B (2$p$). As shown in our recent work,[4] the $Fm\bar{3}m$ symmetry group is ideal for atomic coordinates ($x/a$, $y/b$, $z/c$) but the symmetry of residual electron density (ED) distribution is much lower if non-averaged, accurately measured x-ray data are used in Fourier transform of the structure factors observed. Maxima of residual ED, which are distributed mainly along <110> diagonals of the unit cell, become stronger and form charge stripes directed along one of the selected diagonals at temperatures below the transition temperature to the cage-glass state T* ~ 60 K.[6]

The same crystals demonstrate at low temperatures an essential anisotropy of transverse magnetoresistance whose values change in a good agreement with the stripe directions.[4]

Since cooperative JT effect is determined by the dynamics of light boron atoms, one can suppose that isotope substitutions $^{10}$B–$^{11}$B may affect both properties and crystal structure of dodecaborides. Until now the research was mainly limited by physical properties. The heat capacity and Raman scattering in LuB$_{12}$ have been studied with reference to the crystals of Lu$^{N}$B$_{12}$, N = 10, 11, nat, where 'nat' means natural boron isotope distribution: 18.8% $^{10}$B and 81.2% $^{11}$B. Besides of predictable shift of the Raman spectra to lower frequencies with increasing concentration of $^{11}$B, a broad, low-frequency maximum (boson peak) is observed, which is typical for glasses. The low temperature anomalies of heat capacity combined with the boson peak are interpreted in terms of the transition at $T^* \sim 60$ K into the cage-glass state with static displacements of cations from centers of the B$_{24}$ cavities.[6]

Present work is intended to reveal the structure differences that may affect the crystal properties of Lu$^{N}$B$_{12}$ (N = 10, 11, nat) over the temperature range 88–293 K. Structure studies and measurements of the transport and thermal characteristics (resistance, heat capacity and Seebeck coefficient) were made on the samples prepared from the same single-crystalline discs. The single crystal structures of Lu$^{10}$B$_{12}$ and Lu$^{11}$B$_{12}$ are studied in this work. Data on the structure of Lu$^{nat}$B$_{12}$ at various temperatures are taken both from our previous work (Ref. [4]) and obtained additionally.

## II. EXPERIMENTAL

### A. Crystal growth, sample preparation, charge transport and heat capacity measurements

High-quality single crystals of Lu$^{N}$B$_{12}$, (N = 10, 11 and nat) were grown using the induction zone melting method in an inert gas atmosphere from the preliminarily synthesized LuB$_{12}$ powders.[7] The sample preparation for charge transport and x-ray studies was specified in Ref. [4]. Spherical samples of 0.2–0.3 mm in diameter, most appropriate for the x-ray absorption correction were applied for the crystal structure investigation. The [110] elongated rectangular samples of ≈4×0.2×0.2 mm$^3$ and the bars of ~2×2×0.8 mm$^3$ in size were prepared from the same ingots of Lu$^{N}$B$_{12}$ to investigate the charge transport and heat capacity, correspondingly. For precise measurements of the Seebeck coefficient, a four-terminal scheme was used with step-by-step changes in the temperature gradient on the sample.[8] Commercial PPMS-9 installation (Quantum Design, Inc.) was applied to measure heat capacity. Measurements of resistance were performed in a four-terminal scheme with a direct current commutation (**I** ∥ [110]) at temperatures 1.8–300 K.

## B. X-ray data collection and structure refinement

X-ray data collection was performed using an Xcalibur EOS S2 diffractometer over the temperature range 88–293 K at five temperatures 293, 135, 120, 95, 88 K for $Lu^{N}B_{12}$, N = 10, 11. Two data sets from $Lu^{nat}B_{12}$ at 163 and 201 K were collected additionally to previously collected (Ref. 4) at the same five temperatures. Reproducibility and accuracy of the results is ensured by using original techniques.[9] Data reduction was made using CrysAlisPro (Ref. 10) and ASTRA (Ref. 11) programs. Crystal structures of $Lu^{N}B_{12}$, N = 10, 11, nat, have been refined in the $Fm\bar{3}m$ symmetry group. The independent atomic set consists of two atoms: $Lu_{4a}(0, 0, 0)$ и $B_{48i}(1/2, y, y)$ where subscript characters are Wyckoff positions. Structure refinement and difference Fourier syntheses of ED were made using JANA program.[12] The ED reconstruction by maximal entropy method (MEM) was made by Dysnomia program.[13] The program VESTA (Ref. 14) was used for visualization of the results. Principal data on crystal samples, x-ray experiment and structure refinement details are summarized in Tables SI–SIII (Suppl.).

## III. RESULTS

### A. Dynamic and static components of atomic displacement parameters

Atomic displacement parameters (ADPs) are the refined parameters of the structure model whose values are not tied to the nature of atomic displacements from the lattice points.[15] In the general case, each ADP is a sum of two components. Dynamic component describes thermal and zero temperature vibrations of the atom whereas static component is a temperature-independent mean-square atomic displacement from the lattice point mainly due to boron vacancies and the effect of $^{10}B$-$^{11}B$ isotopic substitution. Harmonic ADPs form a square matrix, the trace of which $<u^2> = u_{obs} = (u_{11} + u_{22} + u_{33})/3$ is known as the observed equivalent value of ADP. The equation:

$$u_{eq}(\text{Lu}) = \frac{h^2}{4\pi^2 * k_B * m_a * T_E}\left(\frac{1}{2} + \frac{1}{\exp\left(\frac{T_E}{T}-1\right)}\right) + \langle u^2 \rangle_{shift(Lu)} \qquad (1)$$

($h$ is the Planck constant; $k_B$ – the Boltzmann constant; $m_a$ – atomic mass; $T_E$ – the characteristic Einstein temperature; $T$ – the temperature of the experiment) links the model equivalent value $u_{eq}(\text{Lu})$ with the Einstein model for independent harmonic oscillators. Weakly bounded Lu atoms well correspond to this model. The second term $\langle u^2 \rangle_{shift(Lu)}$ in the right side of equation accounts a possible static (temperature independent) shift of the Lu atom from the lattice point. Another equation:

$$u_{eq}(\text{B}) = \frac{3*h^2}{4\pi^2 * k_B * m_a * T_D}\left(\frac{1}{4} + \left(\frac{T}{T_D}\right)^2 \int_0^{\frac{T_D}{T}} \frac{y}{\exp(y)-1}dy\right) + \langle u^2 \rangle_{shift(B)} \qquad (2)$$

($T_D$ – the characteristic Debye temperature, $B = {}^{10}B, {}^{11}B, {}^{nat}B$) connects the model equivalent ADP of boron with the Debye model for correlated atomic displacements of the boron framework. The second term $\langle u^2 \rangle_{shift(B)}$ in the right side of Eq. (2) accounts a contribution of a static shift.

The values of $T_E$ ($T_D$) и $<u^2>_{shift}$ are refined using the non-linear least-square (LS) method (Ref. 16) when $u_{eq}$ are approached with $u_{obs}$:

$$\Phi = \sum w[u_{eq} - u_{obs}]^2 \rightarrow \min, \qquad (3)$$

where $w = 1/\sigma^2(u_{obs})$ is a weight of $u_{obs}$. The goodness-of-fit is estimated by the $R1$ factor:

$$R1 = \sum |u_{eq} - u_{obs}| / \sum u_{eq} \qquad (4)$$

The summation in Eq. (3) and (4) is done over the data sets, each of which has been collected at certain temperature to get a single value of $u_{obs}$. Refined values of $T_E$ ($T_D$) and $<u^2>_{shift}$ are then substituted in the right parts of Eq. (1) or (2) to draw the curves presented in Fig. 2(a) or 2(b), respectively.

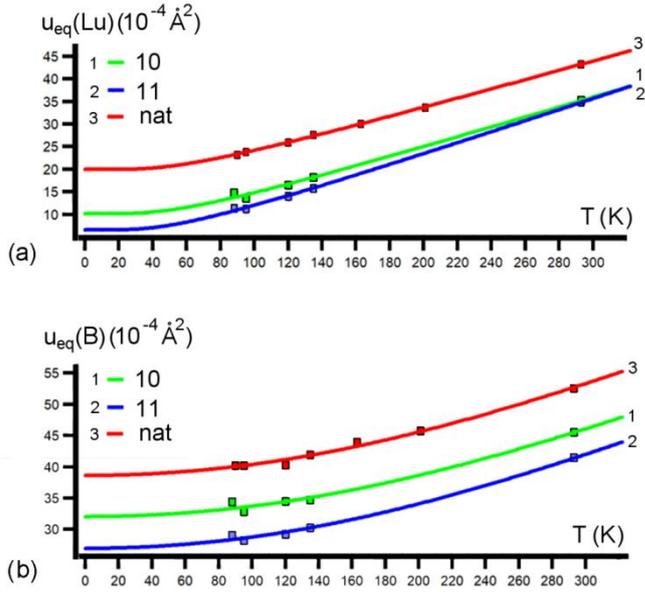

FIG. 2. (Color online) Temperature dependences of $u_{eq}$ in the crystals of $Lu^N B_{12}$, N = 10, 11, nat. The Einstein (a) and Debye (b) models are used respectively for Lu and B atoms. The fit is based on the $u_{obs}$ values marked with squares.

## B. The lower symmetry electron-density distribution in Lu$^N$B$_{12}$

Successful refinement of cubic model does not mean that violations of cubic symmetry are completely excluded in the crystal. Additional structure analysis of Lu$^N$B$_{12}$, N = 10, 11, was performed using accurate, non-averaged x-ray data as it was done before with reference to the Lu$^{nat}$B$_{12}$.[4] Difference Fourier maps and MEM maps are built using the x-ray data collected from the two crystals at five temperatures (Fig. 3). The ED values in a given interval are always divided by VESTA (Ref. 14) into ten levels and a definite color is assigned to each of them. Difference ED values (Δrho) are shown in the layer 0.05 e Å$^{-3}$ < Δrho < 0.8Δrho$_{max}$ changing in color from dark-blue (min) over green to red (max) to provide the most expressive color gradation. The values of rho$_{MEM}$ (full ED) are cut at the level rho$_{max}$ = 0.1% of the peak of rho$_{MEM}$(Lu) to show fine ED gradations in the thin layer. Colors are changing in the reverse sequence, from red (min) to dark-blue (max) to accentuate a distinction between Δrho [Fig. 3(a–d)] and rho$_{MEM}$ [Fig. 3(e, f)].

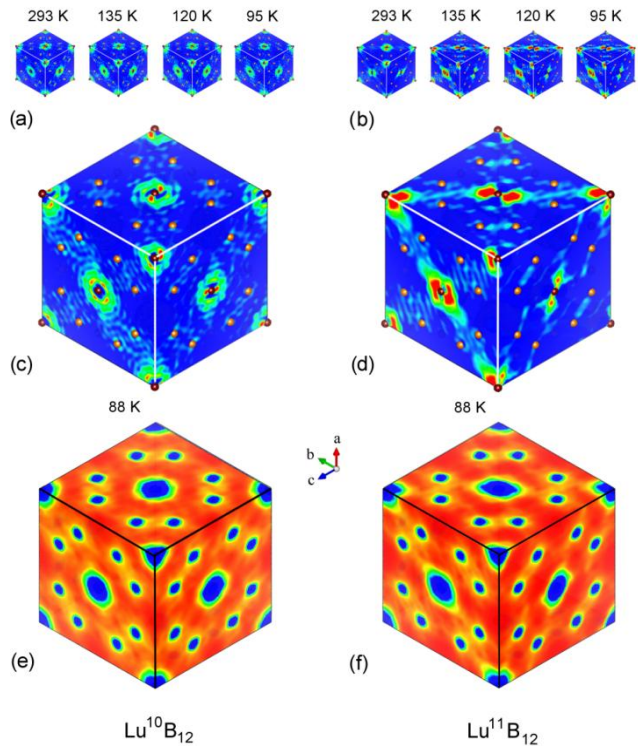

FIG. 3. (Color online) Difference ED maps (a–d) and MEM maps (e, f) are built in sides of the Lu$^N$B$_{12}$ unit cells, N=10, 11. Small dark-red and orange balls (a–d) indicate Lu and B sites, respectively. The maps at 88 K are enlarged to see fine details.

As is evident from Fig. 3, the maps look differently depending on the isotope composition. As to Lu$^{10}$B$_{12}$, the Fourier maps practically do not change with temperature. Two small red spots (residual ED peaks) are seen in the *bc* and *ac* planes near Lu being oriented roughly along the *c* axis whereas small red spots near Lu in the *ab* plane look as if they are connected by a four-fold axis. All this indicates a tetragonal distortion of cubic symmetry. The Fourier maps of Lu$^{10}$B$_{12}$ and Lu$^{11}$B$_{12}$ are similar at 293 K but differences become stronger with temperature decreasing. Large, [110]-oriented red spots are clearly seen near Lu in case of Lu$^{11}$B$_{12}$. One can distinguish three side diagonals connected by the [1-11] axis, which is perpendicular to the plane of the figure. In other words, it is rather a trigonal distortion than tetragonal one. Some signs of the axis 4 ∥ [001] can be seen, however, in the *ab* plane judging from the residual ED distribution near the central Lu site.

### C. Temperature dependences of the unit-cell values

Lower symmetry ED distribution points to probable lattice distortions. Non-averaged, linear (Fig. 4) and angular (Fig. 5) unit-cell values of Lu$^N$B$_{12}$, (N = 10, 11, nat) are defined at several temperatures from x-ray data.

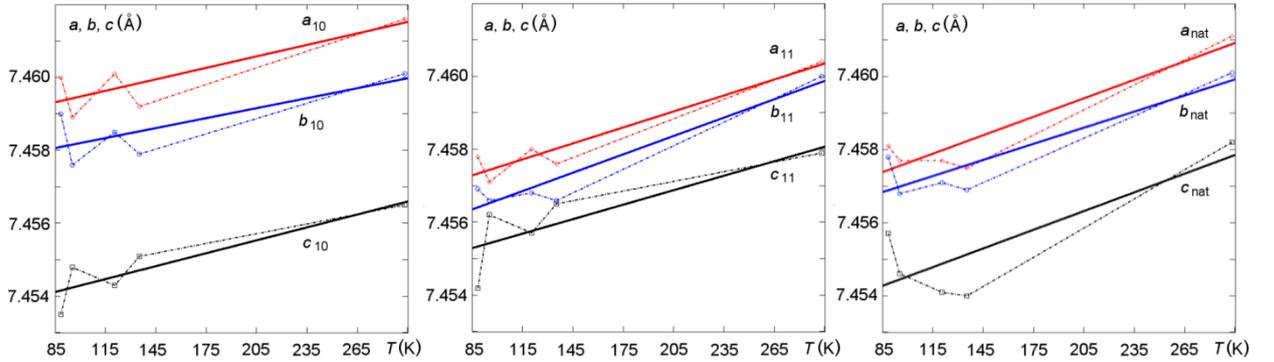

FIG. 4. (Color online) Temperature dependence of the Lu$^N$B$_{12}$ lattice periods over the temperature range 88–293 K. Experimental values are connected by dash-dot lines; solid lines are linear approximation. Standard uncertainties do not exceed 0.0002 Å.

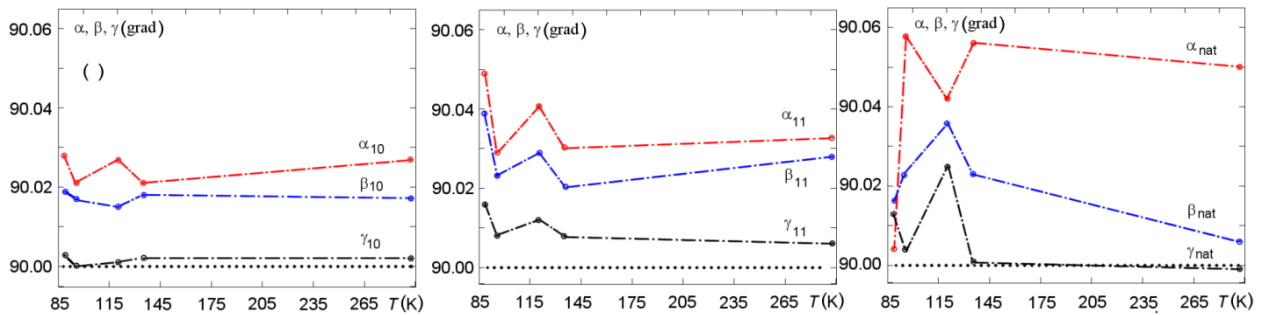

FIG. 5. (Color online) Temperature dependence of the Lu$^N$B$_{12}$ unit-cell angles over the temperature range 88–293 K. Experimental values are connected by dash-dot lines. Standard uncertainties do not exceed 0.001°.

As can be seen from Fig. 4 and 5, tetragonal deformation of the crystal lattice prevails in Lu$^{10}$B$_{12}$ whereas the lattice of Lu$^{11}$B$_{12}$ is distorted mainly by trigonal type. Changes of the lattice periods are nonmonotonic within 0.001 Å and qualitatively similar in different crystals what is difficult to explain by a random spread only. Angular deviations from 90° also obey definite regularities. They are maximal in the lattice of Lu$^{nat}$B$_{12}$ and minimal in the lattice of Lu$^{10}$B$_{10}$ in combination with more pronounced tetragonal deformation of the periods. Less deformed angle is always between two longer periods and vice versa. For instance, the angle γ between 'long' periods *a*, *b* is usually less deviated from 90° than two other angles.

### D. Resistivity, specific heat and Seebeck coefficient in Lu$^N$B$_{12}$

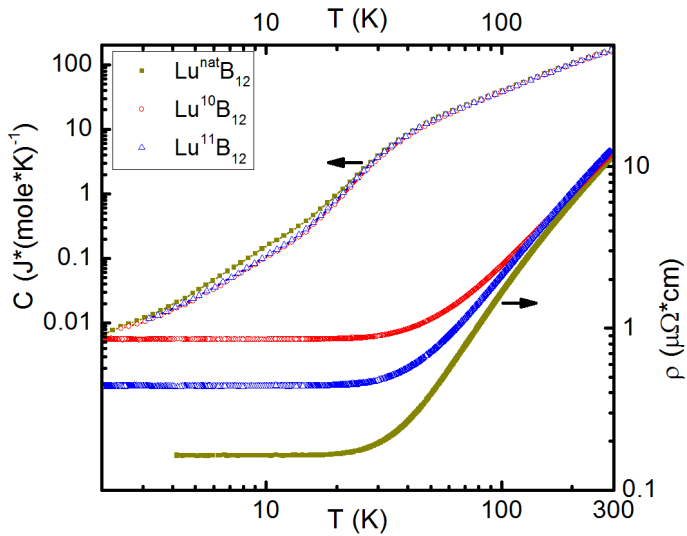

FIG. 6. (Color online) Temperature dependencies of the specific heat and resistivity of the crystals Lu$^N$B$_{12}$, N = 10, 11 and nat.

Temperature dependent resistivity ρ and specific heat C of Lu$^N$B$_{12}$ (N = 10, 11, nat) are shown in Fig. 6. The resistivity decreases in the interval 2–300 K with temperature lowering and reaches residual value $\rho_0$ below 20 K. For Lu$^{nat}$B$_{12}$, the $\rho(T)/\rho_0$ ratio (RRR) reaches a maximum (RRR ≈ 70) with minimal resistivity $\rho_0$ ~ 0.15 µOhm·cm while $\rho_0$ is essentially higher for isotopically pure crystals (see Fig. 6 and Table II below).

For Lu$^N$B$_{12}$ (N = 10, 11, nat), temperature dependences of the specific heat C(T) at a constant pressure and intermediate temperatures T > 30 K almost coincide with each other in the log–log plot (Fig. 6) whereas noticeable differences in C(T) are observed at low temperatures.

The largest $C(T)$ values in the range of 2–20 K were detected for the Lu$^{nat}$B$_{12}$ assuming effects in the specific heat of this compound of random isotope substitutions and boron vacancies. A 'step-like' anomaly on the $C(T)$ dependence in the range 20–30 K for Lu$^{nat}$B$_{12}$ (Fig. 6) was observed and discussed in terms of the Einstein type contribution to the specific heat from quasi-local vibrations of rare earth ions embedded in the large-size B$_{24}$ cavities.[6,17,18] According to the data presented in Fig. 6, boron isotope substitution affects insignificantly the behavior of the Einstein component in the specific heat of Lu$^N$B$_{12}$ (N = 10, 11, nat), what confirms the loosely bound state of the Lu$^{3+}$ ions. Temperature dependences of the Seebeck coefficient $S(T)$ (Fig. 7) demonstrate more or less pronounced negative minima at intermediate temperatures 30–150 K.

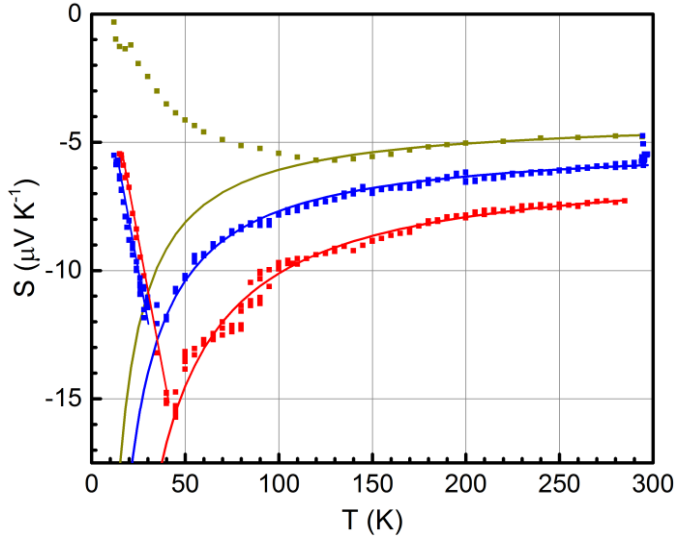

FIG. 7. (Color online) The temperature dependencies of the Seebeck coefficient of the crystals Lu$^N$B$_{12}$, N = 10, 11 and nat. Solid lines show the data approximation by Mott dependence and phonon drag thermopower [see Eq. (10) and text].

The amplitude of the anomaly, which is observed for all the dodecaborides under study, is the largest one in Lu$^{10}$B$_{12}$ and the smallest one in Lu$^{nat}$B$_{12}$ (Fig. 7). At low temperatures the Seebeck coefficient changes linearly, and the largest slope of this Mott (diffusive) thermopower is detected in the isotopically pure lutetium dodecaborides.

## IV. DISCUSSION

### A. Crystal structure

Refined characteristic Einstein and Debye temperatures and temperature independent ('constant') components of equivalent ADPs, $\langle u^2 \rangle_c = \langle u^2 \rangle_{shift} + \langle u^2 \rangle_{zero}$, are shown in Table I for each of Lu$^N$B$_{12}$, N = 10, 11, nat.

TABLE I. The Einstein and Debye characteristic temperatures and temperature independent components of equivalent ADPs refined by LS method with goodness-of-fit $R1$.

|  | Lu$^{10}$B$_{12}$ | Lu$^{11}$B$_{12}$ | Lu$^{nat}$B$_{12}$ |
|---|---|---|---|
| $T_E$ (K) | 157(3) | 149(1) | 162(1) |
| $<u^2>_c$(Lu) (Å$^2$) | 0.00101(8) | 0.00066(4) | 0.00200(2) |
| $R1$,% | 2.26 | 1.33 | 0.47 |
| $T_D$ (K) | 1126(31) | 1061(19) | 1077(18) |
| $<u^2>_c$(B) (Å$^2$) | 0.00320(13) | 0.00270(8) | 0.00386(8) |
| $R1$ (%) | 1.25 | 0.99 | 0.69 |

The Debye temperatures coincide within one standard uncertainty ($\sigma$) so they may be averaged to $T_D$=1088(44) K. The Einstein temperatures differ not more than by 7$\sigma$ with average value of $T_E$=155.8(7) K. Smaller values of $u_{eq}$(Lu) and $u_{eq}$(B) in Lu$^{11}$B$_{12}$ as compared to Lu$^{10}$B$_{12}$ (Fig. 2) can be presumably explained by higher perfection of the first crystal and heavier $^{11}$B atoms as to boron. From three crystals of different isotopic composition, a secondary extinction is less in Lu$^{10}$B$_{12}$, high in Lu$^{nat}$B$_{12}$ and maximal in Lu$^{11}$B$_{12}$. One may expect their diffraction quality to grow in the same order.

ADPs summarize mean-square displacements from zero vibrations $<u^2>_{zero}$, temperature dependent thermal vibrations $<u^2(T)>$ and static shifts $<u^2>_{shift}$. The curves in Fig. 2 are built for three crystals with very close Debye (Einstein) temperatures, so the $u_{eq}(T)$ are notably differ over the difference in their temperature independent components. As is seen from Table I, these ADP components $<u^2>_c = <u^2>_{zero} + <u^2>_{shift}$ are maximal in Lu$^{nat}$B$_{12}$ both for Lu and B atoms. Static distortions of boron polyhedrons are combined with static shifts of Lu atoms from the inversion centers what can be explained by the JT instability of the B$_{12}$ clusters and the disorder in $^{10}$B–$^{11}$B substitution in the crystal with natural boron.

The first attempt to analyze the residual ED distribution in Lu$^{nat}$B$_{12}$ without relying on the symmetry of the structural model was made in (Ref. 4). In present work, the same approach is applied to two other crystals of Lu$^N$B$_{12}$ different in isotope composition. In addition, non-averaged unit-cell values of the three Lu$^N$B$_{12}$ crystals, N = 10, 11, nat, are first analyzed here. The lattice deviations from cubic symmetry by mainly tetragonal (Lu$^{10}$B$_{12}$), mainly trigonal (Lu$^{11}$B$_{12}$) or mixed (Lu$^{nat}$B$_{12}$) type are repeated in character of the ED distribution. This observation deserves attention since unit-cell values depend on accurate centering of the x-ray reflections whereas quality of the Fourier syntheses is mainly determined by integral intensities of the reflections. The centering procedure does not operate with integral intensities, and there is no need in precise centering of each reflection to provide accurate measurements of integral intensities, particularly in modern coordinate diffractometers.

### B. Crystal properties

#### 1. Resistivity

The resistivity changes with temperature $\Delta\rho(T)$ are analyzed here in terms of the Einstein formula:[19]

$$\Delta\rho = \rho - \rho_0 = \rho_E = \frac{A}{T\left(e^{\frac{T_E}{T}} - 1\right)\left(1 - e^{-\frac{T_E}{T}}\right)}, \quad (5)$$

which is expected to be valid at $T < T^* \sim 60$ K in the cage-glass state of the $Lu^N B_{12}$ crystals with strong electron-phonon scattering on the quasi-local vibrations of the $Lu^{3+}$ ions. Figure 8(a) shows a fit of the resistivity data by Eq. (5), which allows estimating $T_E = 162–170$ K (see Table II).

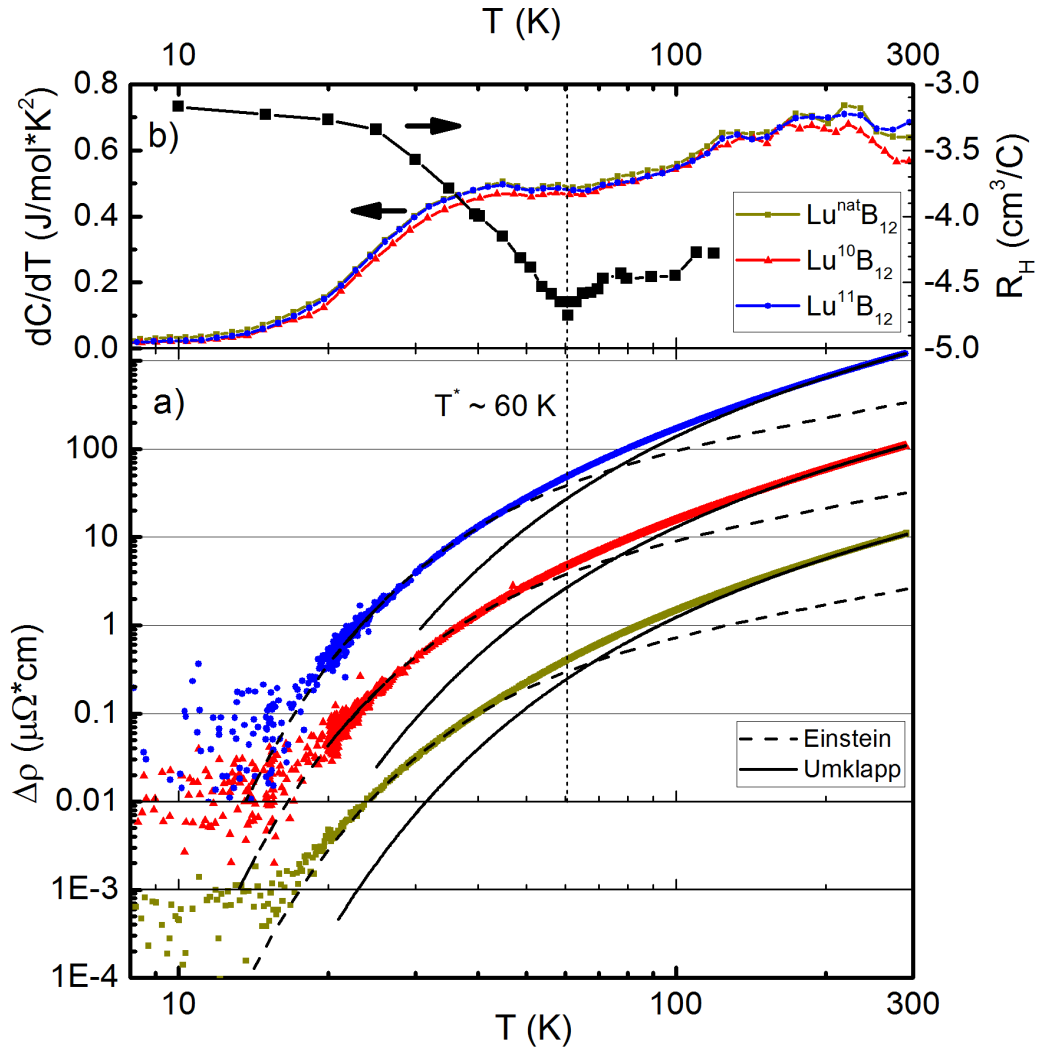

FIG. 8. (Color online) (a) Temperature dependent component of the resistivity curves $\Delta\rho(T)$ (curves are shifted for convenience). Fitting results from Eq. (5) and (6) are shown by dashed and solid lines, respectively. (b) Temperature dependences of the derivative of specific heat $dC/dT$ and the Hall coefficient $R_H(T)$.

TABLE II. Parameters detected from the resistivity data analysis [see Eq. (5), (6) and text].

|  | $A$ (μΩ·K) | $T_E$ (K) | $B_0$ (K$^{-1}$) | $T_0$ (K) | $\rho_0$ (μOhm cm) |
|---|---|---|---|---|---|
| Lu$^{10}$B$_{12}$ | 297.0 | 162.8 | 0.066 | 163 | 0.85 |
| Lu$^{11}$B$_{12}$ | 344.6 | 169.7 | 0.076 | 170 | 0.43 |
| Lu$^{nat}$B$_{12}$ | 257.8 | 168.5 | 0.066 | 169 | 0.15 |

Additionally, Figure 8(b) performs temperature dependences of the derivative of specific heat $dC/dT=f(T)$ and the Hall coefficient $R_H(T)$ demonstrating anomalies at $T^* \sim 60$ K. It can be seen that Umklapp processes dominate above $T^*$ in the charge carrier scattering of LuB$_{12}$ and relation

$$\Delta\rho = \rho_U = B_0 T e^{-\frac{T_0}{T}} \tag{6}$$

may be considered as a good approximation for resistivity in the temperature range 100–300 K. It is worth noting that $T_0$ parameter in Eq. (6) detected from these fits is close to the Einstein temperatures $T_E$ (see Table II). At the same time the crossover between these two regimes described by Eq. (5) and (6) coincides very well with the $T^*$ transition area (see Fig. 8). Taking into account that $T_0 \sim v_s \mathbf{q}/k_B$, where $v_s$ is sound velocity, $\mathbf{q}$ is the wave vector of the Umklapp phonon and $k_B$ is the Boltzmann constant, one can consider the quasi-local modes of Lu-ions as vibrations prevailing in the Umklapp processes in LuB$_{12}$.

*2. Specific heat*

To analyze the contributions to the specific heat of the Lu$^N$B$_{12}$ samples, we used an approach similar to that previously applied in (Ref. 6). Contributions from B and Lu atoms in the vibrational heat capacity are considered in terms of the Debye [Eq. (7)] and Einstein [Eq. (8)] models, respectively:

$$\frac{C_D}{T^3} = \frac{9rR}{T_D^3} \int_0^{T_D/T} \frac{x^4 e^{-x} dx}{(1-e^{-x})^2} \tag{7}$$

$$\frac{C_E}{T^3} = \frac{3R}{T_E^3}\left(\frac{T_E}{T}\right)^5 \frac{e^{-\frac{T_E}{T}}}{(1-e^{-\frac{T_E}{T}})^2}, \tag{8}$$

where $R$ is gas constant and $r = 12$. The electronic specific heat $C_{el} = \gamma T$ with $\gamma \approx 3$ mJ/(mol K$^2$) and the Debye contribution with $T_D$ detected above from $u_{eq}$(B) approximation by Eq. (2) were applied here to calculate the difference $C_{ph}= C - C_{el} - C_D = C_E(T) + C_{Sch(i)}(T)$. Following to (Ref. 6) we used additionally two Schottky terms $C_{Sch(i)}$ – two-level-systems TLS$_1$ and TLS$_2$ to approximate low temperature anomalies of the specific heat. It has been argued in (Ref. 6, 20) that these two Schottky components:

$$\frac{C_{Sch(i)}}{T^3} = \frac{RN_i}{T^3}\left(\frac{\Delta E_i}{T}\right)^2 \frac{e^{\frac{\Delta E_i}{T}}}{(e^{\frac{\Delta E_i}{T}}+1)^2} \quad (9)$$

($i$ = 1, 2 and $N_i$ is the concentration of two-level-systems) are necessary to describe the effect of boron vacancies (TLS$_2$ contribution) and divacancies (TLS$_1$) in the specific heat of $R$B$_{12}$.

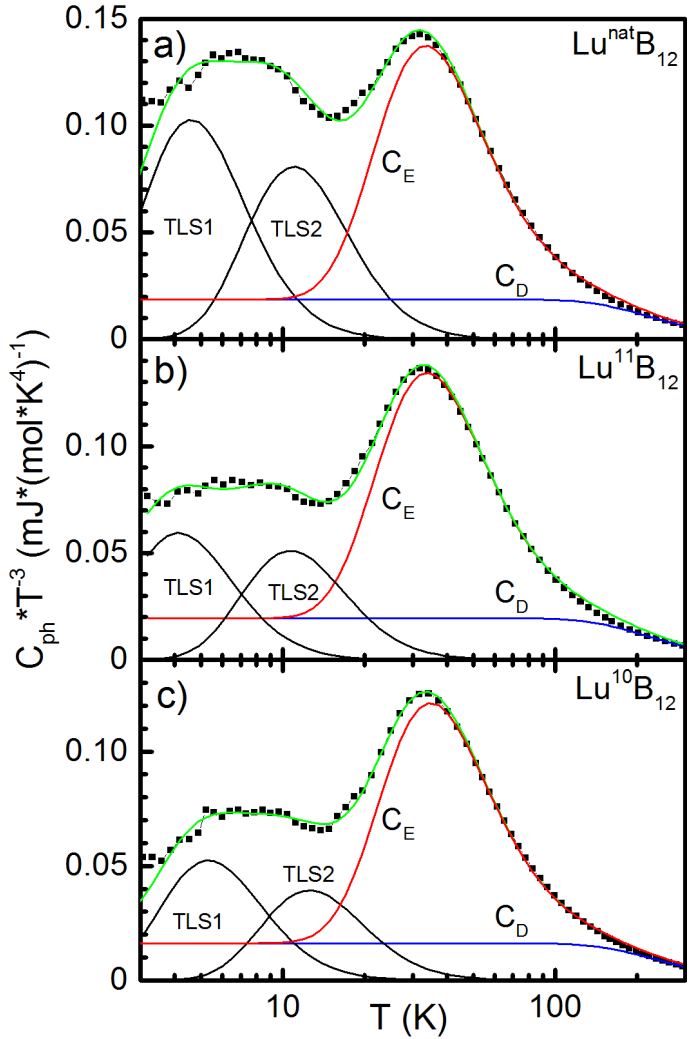

FIG. 9. (Color online) Separation of the low temperature vibrational contribution $C_{ph}/T^3$ to the specific heat of Lu$^N$B$_{12}$ into Debye ($C_D$), Einstein ($C_E$) and two Schottky (TLS$_1$ and TLS$_2$) components (see text).

Indeed, in view of a weak coupling of the rare-earth ions in the boron network in combination with a significant number of boron vacancies and other intrinsic defects in the UB$_{12}$ type structure,[21] the formation of double-well potentials (DWP) should be expected at displacements of the Lu$^{3+}$ ions from central positions in B$_{24}$ cubooctahedra. All the above mentioned specific heat contributions are shown in Fig. 9 in the $C/T^3$ plot together with our experimental data.

According to the approach developed in (Ref. 6, 20), the energy $\Delta E_2/k_B$ = 54–64 K [see Eq. (9) and Table III below], which is deduced from the analysis developed here, should be attributed to the barrier height in the DWP. Moreover, the normalized concentration $N_2$ = 0.047–0.08 of TLS$_2$ (see Table III) corresponds to the number of lutetium ions displaced from the central positions in B$_{24}$ cells. The result is in a good agreement with the concentration 0.036 deduced from the EXAFS measurements of LuB$_{12}$ powders at low temperatures.[22] Since boron vacancies seem be the main source to the formation of intrinsic defects in the structure of $R$B$_{12}$ and each boron vacancy ensures the displacement of two neighboring rare-earth ions from the center of the B$_{24}$ octahedron, the number of boron vacancies can be estimated as $n_v = N_2/2$ = 2.4–4%. Both the concentration of boron divacancies $N_1 \equiv n_d$ = 0.0032–0.0075 and the barrier height $\Delta E_{TLS1}/k_B$= 21–27 K are also detected naturally for Lu$^N$B$_{12}$ crystals studied (see Table III).

It is interesting to note here that in the case of random distribution of boron vacancies in $R$B$_{12}$ a simple combinatorial relation $n_d = n_v(1 - (1 - n_v)^z)$ ($z = 4$ is the coordination number in the boron lattice) may be used to estimate the expected divacancy concentration. We consider the calculated value to be in a good agreement with $n_d$ deduced from the above developed heat capacity analysis. It is worth noting also that the barrier height $\Delta E_2/k_B$ in the double-well potential for Lu$^N$B$_{12}$ is close to the cage-glass transition temperature $T^*$ = 54–65 K.[6, 23] This allows one to conclude in favor of the simple scenario of the order-disorder transition in the rare earth ion sublattice in terms of 'freezing' of Lu$^{3+}$ ions at different positions of DWP minima induced by the random distribution of boron vacancies at temperature decreasing below $T^*$.

TABLE III. Barrier heights $\Delta E_1$ and $\Delta E_2$ in the double-well potential, the concentration of TLS$_1$ ($N_1$) and TLS$_2$ ($N_2$), Einstein temperature $T_E$ as detected from the heat capacity analysis [Eq. (7)–(9)], $T^*$ is the cage-glass transition temperature.[6, 23]

| Lu$^N$B$_{12}$ | $\Delta E_{TLS1}$ (K) | $N_2$(B)/2 = $n_v$ | $\Delta E_{TLS2}$ (K) | $N_1$(B) = $n_d$ | $T_E$ (K) | $T^*$ (K) |
|---|---|---|---|---|---|---|
| Lu$^{10}$B$_{12}$ | 63.5 | 0.058/2 → 3% | 27 | 0.006 | 171.9 | 65 |
| Lu$^{11}$B$_{12}$ | 54.1 | 0.047/2 → 2.4% | 20.9 | 0.0032 | 166.8 | 58 |
| Lu$^{nat}$B$_{12}$ | 55.4 | 0.08/2 → 4% | 23.3 | 0.0075 | 165 | 54 |

*3. Seebeck coefficient*

When discussing the Seebeck coefficient behavior (Fig. 7), it is worth noting that negative $S(T)$ minimum is typical feature for metals with electron conduction (see, for example, Ref. 24) and it appears as a crossover from phonon-drag thermopower with the dependence $S_g \sim 1/T$ at higher temperatures to a linear diffusive low temperature component $S = BT$. As a result, these two limits of the $S(T)$ behavior may be described as:[24]

$$S(T) = BT + K/T + S_0. \qquad (10)$$

These two parts of thermopower are approximated by solid lines in Fig. 7 and fitting parameters are collected in Table IV. Taking into account that electron-phonon scattering on quasi-local vibrations of $Lu^{3+}$ ions dominates over intermediate temperature range, it is natural to expect the Einstein modes to be determinants for thermopower of the phonon dragging, too. At the same time, it does not seem possible in frames of existing models to relate directly the coefficient $K$ in Eq. (10) to characteristics of the Einstein phonons.

TABLE IV. Parameters deduced in the analysis of the Seebeck coefficient temperature dependences [see Fig. 7 and Eq. (10)].

| $Lu^N B_{12}$ | $B$ (µV/K$^2$) | $K$ (µV) | $S_0$ (µV/K) |
|---|---|---|---|
| $Lu^{10}B_{12}$ | 0.38 | -439.4 | -5.72 |
| $Lu^{11}B_{12}$ | 0.42 | -270.6 | -4.98 |
| $Lu^{nat}B_{12}$ | 0.084 | -204.9 | -4.02 |

### C. The 'isotope composition – structure – properties' relationship

Turning to a general discussion, we should highlight the high quality of the single crystals under investigation. It is worth noting that optical conductivity of $Lu^N B_{12}$ (N = 10, 11, nat) was studied in (Ref. 25) on the same crystals. It was determined that about 70% of the charge carriers in these dodecaborides were involved into collective modes and only 30% of the conduction electrons were the Drude-like free carriers. Moreover, it was suggested in (Ref. 25) that the origin of these non-equilibrium electrons was connected with cooperative dynamics of the Jahn–Teller active $B_{12}$ molecules producing quasi-local vibrations (rattling modes) of caged lutetium ions (see also Ref. 4). The coupling of $Lu^{3+}$ rattling motions with the charge carriers of conduction band was proposed to be the reason of strongly damped character of the excitations. Considering close values of the parameters characterizing collective modes in the $Lu^N B_{12}$ crystals,[25] it is natural to expect that differences boil down to moderate changes of the Debye and Einstein temperatures $T_D$ = 1060–1130 K and $T_E$ = 150–170 K (Tables I–III) as follows from the analysis of experimental results (Figures 2 and 7–9). Maximal value of $T_D$ = 1130 K is naturally related to the light isotope $^{10}B$ whose upper limit of phonon spectrum is remarkably shifted as compared to $Lu^{nat}B_{12}$ and $Lu^{11}B_{12}$.[5, 7]

The isotope different crystals of Lu$^N$B$_{12}$ give first information on probable structure prerequisites for the charge stripe formation. In contrast to Lu$^{nat}$B$_{12}$, in which dynamical charge stripes are observed at $T = 50$ K $< T^*$,[4] and unlike Lu$^{nat}$B$_{12}$ and Lu$^{11}$B$_{12}$ where traces of stripes are reliably recorded at temperatures 88–135 K, no trace of stripes is observed in the difference Fourier maps of Lu$^{10}$B$_{12}$. Instead of this, residual ED forms rings round the Lu sites and weak interstitial spots (Fig. 3). One can assume probable relation between the presence of $^{11}$B in the $R$B$_{12}$ composition and trigonal structure distortions, on the one hand, and the emergence of the dynamic charge stripes in a certain dodecaboride crystal, on the other.

The least values of the temperature independent mean-square displacements $<u^2>_c$ are registered for Lu$^{11}$B$_{12}$ with moderate structure distortions. On the contrary, the highest values of $<u^2>_c$ correspond to Lu$^{nat}$B$_{12}$ with largest structure distortions (see Figures 2, 4, 5 and Table I). The mixed isotope composition is a native defect affecting also as a factor, which increases the low temperature specific heat of Lu$^{nat}$B$_{12}$, see Figures 6 and 9. It looks like an obvious contradiction when strongly disordered crystal shows low values of the charge transport parameters $\rho_0$, $S_0$ and $B$, see Tables II and IV. One can suppose, however, that the values of $\rho_0$ и $S_0$ are lowered in these 'defective' crystals of Lu$^{nat}$B$_{12}$ owing to high-conductive percolation channels (dynamic charge stripes), which leads to a shorting out of charge transfer in the matrix. Taking into account that the coefficient $B$ in Eq. (10) is defined by the Mott formula for the electronic thermopower:

$$S_d(T) = \pi^2 k_B^2 T/(3e)(\partial \ln g(E_F)/\partial E) \equiv BT, \quad (11)$$

where $g(E_F)$ is a derivative of the electron density of states at the Fermi level $E_F$, one can explain the fivefold difference (Table IV) in $B$-values between Lu$^{nat}$B$_{12}$ and isotopically pure Lu$^N$B$_{12}$ whose charge stripes (if exist) apparently do not form 'infinite' filamentary conductive channels. In this scenario significant effective decrease of the $B$ parameter in Lu$^{nat}$B$_{12}$ can be explained by the nanoscale heterogeneity due to strongly conductive channels and despite the fact that electron properties of the higher boride frameworks are practically equal for each of Lu$^N$B$_{12}$. Note that values of resistivity and thermopower are intermediate for the least defective Lu$^{11}$B$_{12}$ crystal whose charge stripes are weak as compared with those in Lu$^{nat}$B$_{12}$. As for Lu$^{10}$B$_{12}$ where no additional filamentary structure of high-conductive channels has been observed, the measured values of $\rho_0$ и $S_0$ are typical for the dodecaboride matrix and are even increased due to heterogeneous but not filamentary distribution of the residual ED.

The above new data are not enough to build a full system of cause-and-effect relations between parameters of the JT lattice distortion of the boron lattice and a certain probability of the dynamic charge stripe formation. It seems that additional channels of charge transport can be formed in $R$B$_{12}$ in case of noticeable trigonal distortions of the cubic lattice in combination with optimal number of structure defects, which are enough to form selected channels (stripes) but do not exceed the value, above which they break into short chains. Note that measurements of heat capacity and charge transport (resistivity, Seebeck coefficient) are complementary experiments. In case the dynamical charge stripes are defined from the structure data, their consideration together with the above values can give new data for analysis. Heat capacity, as a bulk material property, can be used at low temperatures to control quality of materials (concentration and categories of defects) whereas transport parameters are integral characteristics of the heterogeneous medium including both filamentary structure of conductive channels and basic metallic matrix of the compound.

## V. CONCLUSIONS

Structure differences of isotopically different crystals of $Lu^NB_{12}$ (N = 10, 11, nat) have been reliably determined for the first time and their impact on thermal and charge transport characteristics has been discussed. Three types of the structure distortions are observed in the dodecaborides studied here. 1. Local atomic disordering (defects), which is particularly expressed in $Lu^{nat}B_{12}$ judging from maximal values of both the temperature independent components of ADPs and the Schottky anomalies in the heat capacity. 2. The long-range JT disordering of the crystal lattice, which can be trigonal or tetragonal. 3. The 'medium-range' distortion of the ED distribution, which reveals itself in some cases (trigonal lattice distortion plus defects) as dynamic charge stripes. In case of remarkable trigonal lattice distortion supplemented with strong static displacements of the atoms, the effect goes into a 'practical' stage: the conductivity of the heterogeneous media increases as one can see in $Lu^{nat}B_{12}$ for which the contradiction is removed between the strong local disordering and high conductivity. The defects are supposedly used as the centers of pinning facilitating formation of additional conductive channels. How can defects (something random) ensure ordering? The answer is more than a hypothesis – due to the cooperative dynamic JT effect when the $B_{12}$ polyhedra are consistently distorted.

## ACKNOWLEDGEMENTS

This work was supported by the Russian Ministry of Science and Higher Education within the State assignment FSRC 'Crystallography and Photonics' RAS of X-ray diagnostics, by the Russian Foundation for Basic Research [grant no. 16-02-00171] in part of the crystal structure and property research, and by the Program of Fundamental Research of the Presidium of the Russian Academy of Sciences 'Fundamental Problems of High-Temperature Superconductivity' in part of resistance measurements and analysis. The X-ray data were collected using the equipment of the Shared Research Center FSRC 'Crystallography and Photonics' RAS whose work was partially supported by the Russian Ministry of Science and Higher Education.

[1] K. Flachbart, P. Alekseev, G. Grechnev, N. Shitsevalova, K. Siemensmeyer, N. Sluchanko, and O. Zogal, in *Rare Earths: Research and Applications* ed K N Delfrey (Hauppauge, NY: Nova Science) pp 79–125 (2008).

[2] D. A. Voronovich, A. A. Taran, N. Yu. Shitsevalova, G. V. Levchenko, and V.B. Filipov, Funct. Mater. **21**, 266 (2014). http://dx.doi.org/10.15407/fm21.03.266

[3] N. Sluchanko, A. Bogach, N. Bolotina, V. Glushkov, S. Demishev, A. Dudka, V. Krasnorussky, O. Khrykina, K. Krasikov, V. Mironov, V. Filippov, and N. Shitsevalova, Phys. Rev. B **97**, 035150 (2018). https://doi.org/10.1103/PhysRevB.97.035150

[4] N. B. Bolotina, A. P. Dudka, O. N. Khrykina, V. N. Krasnorussky, N. Yu. Shitsevalova, V. B. Filipov, and N. E. Sluchanko, J. Phys.: Condens. Matter **30**, 265402 (2018). https://doi.org/10.1088/1361-648X/aac742

[5] H. Werheit, Yu. Paderno, V. Filippov, V. Paderno, A. Pietraszko, M. Armbrüster, and U. Schwarz, J. Solid State Chem. **179**, 2761 (2006). https://doi.org/10.1016/j.jssc.2005.11.034

[6] N. E. Sluchanko, A. N. Azarevich, A. V. Bogach, I. I. Vlasov, V. V. Glushkov, S. V. Demishev, A. A. Maksimov, I. I. Tartakovskii, E. V. Filatov, K. Flachbart, S. Gabani, V. B. Filippov, N. Yu. Shitsevalova, and V. V. Moshchalkov, J. Exp. Theor. Phys. **113**, 468 (2011). https://doi.org/10.1134/S1063776111080103

[7] H. Werheit, V. Filipov, K. Shirai, H. Dekura, N. Shitsevalova, U. Schwarz. and M. Armbruster, J. Phys.: Condens. Matter **23**, 065403 (2011). https://doi.org/10.1088/0953-8984/23/6/065403

[8] S. V. Demishev, M. V. Kondrin, V. V. Glushkov, N. E. Sluchanko, and N. A. Samarin, J. Exp. Theor. Phys. 86, 182 (1998). https://doi.org/10.1134/1.558482

[9] A. P Dudka, Crystallography Reports **63**, 1051 (2018). DOI: 10.1134/S1063774518050085

[10] Rigaku Oxford Diffraction, CrysAlisPro Software system, version 1.171.39.15e, Rigaku Corporation, Oxford, UK (2016).

[11] A. Dudka, J. Appl. Crystallogr. **40**, 602 (2007). http://dx.doi.org/10.1107/S0021889807010618

[12] V. Petricek, M. Dusek, and L. Palatinus, Z. Kristallogr. **229**, 345 (2014). https://doi.org/10.1515/zkri-2014-1737

[13] K. Momma, T. Ikeda, A. A. Belik, and F. Izumi, Powder Diffr. **28**, 184 (2013). https://doi.org/10.1017/S088571561300002X

[14] K. Momma and F. Izumi, J. Appl. Crystallogr. **44**, 1272 (2011). https://doi.org/10.1107/S0021889811038970

[15]K. N. Trueblood, H-B. Bürgi, H. Burzlaff, J. D. Dunitz, C. M. Gramaccioli, H. H. Schulz, U. Shmueli, and S. C. Abrahams, Acta Crystallogr. Sect. A **52**, 770 (1996). http://dx.doi.org/10.1107/S0108767396005697

[16]A. P. Dudka, N. B. Bolotina, O. N. Khrykina, J. Appl. Crystallogr. (to be published).

[17]A. Czopnik, N. Shitsevalova, A. Krivchikov, V. Pluzhnikov, Y. Paderno, and Y. Onuki, J. Solid State Chem. **177**, 507 (2004). https://doi.org/10.1016/j.jssc.2003.04.003

[18]A. Czopnik, N. Shitsevalova, V. Pluzhnikov, A. Krivchikov, Yu. Paderno, and Y. Onuki, J. Phys.: Condens. Matter **17**, 5971 (2005). https://doi.org/10.1088/0953-8984/17/38/003

[19]J. R. Cooper, Phys. Rev. B **9** 2778 (1974). https://doi.org/10.1103/PhysRevB.9.2778

[20]N. E. Sluchanko, A. N. Azarevich, S. Yu. Gavrilkin, V. V. Glushkov, S. V. Demishev, N. Yu. Shitsevalova, and V. B. Filippov, JETP Letters **98**, 578 (2013). https://doi.org/10.1134/S0021364013220141

[21]Z. Fojud, P. Herzig, O. J. Zogal, A. Pietraszko, A. Dukhnenko, S. Jurga, and N. Shitsevalova, Phys. Rev. B **75**, 184102 (2007). https://doi.org/10.1103/PhysRevB.75.184102

[22]A. P. Menushenkov, A. A. Yaroslavtsev, I. A. Zaluzhnyy, A. V. Kuznetsov, R. V. Chernikov, N. Y. Shitsevalova, and V. B. Filippov, JETP Letters **98**, 165 (2013). https://doi.org/10.1134/S002136401316011X

[23]N. E. Sluchanko, A. N. Azarevich, A. V. Bogach, V. V. Glushkov, S. V. Demishev, A. V. Kuznetsov, K. S. Lyubshov, V. B. Filippov, and N. Yu. Shitsevalova, J. Exp. Theor. Phys. **111**, 279 (2010). https://doi.org/10.1134/S1063776110080212

[24]F. J. Blatt, P. A. Schroeder, C. L. Foiles, D. Greig. Thermoelectric Power of Metals. Publisher: Plenum Publishing Corporation, New York (1976).
DOI: 10.1007/978-1-4613-4268-7

[25]B. P. Gorshunov, E. S. Zhukova, G. A. Komandin , V. I. Torgashev, A. V. Muratov, Yu. A. Aleshchenko, S. V. Demishev, N. Yu. Shitsevalova, V. B. Filipov, and N. E. Sluchanko, JETP Letters **107,** 100 (2018). https://doi.org/10.1134/S0021364018020029